\documentclass[aps,nofootinbib,superscriptaddress,preprint]{revtex4-1}

\pdfoutput=1
\usepackage{graphicx}
\usepackage{amsfonts}
\usepackage{amssymb}
\usepackage{amsmath}
\usepackage{mathtools}
\usepackage{txfonts}
\usepackage{lipsum}
\usepackage{color}
\usepackage{wasysym}
\usepackage{bbold}
\usepackage[colorlinks=true, urlcolor=blue, linkcolor=blue,
citecolor=blue, hyperindex=true, linktocpage=true]{hyperref}

\DeclareMathOperator{\Tr}{Tr}

\usepackage{letltxmacro}
\LetLtxMacro{\oldsqrt}{\sqrt}
\renewcommand{\sqrt}[2][\mkern8mu]{\mkern-6mu\mathop{}\oldsqrt[#1]{#2}}
\makeatletter
\def\blfootnote{\xdef\@thefnmark{}\@footnotetext}
\makeatother

\begin{document}
\title{Detecting quantum critical points in the $t-t'$ Fermi-Hubbard model via
complex network theory}

\author{Andrey A. Bagrov}
\email{andrey.bagrov@physics.uu.se}
\affiliation{Department of Physics and Astronomy, Uppsala University, Box 516, SE-75120 Uppsala,
Sweden}
\affiliation{Radboud University, Institute for Molecules and Materials, 6525AJ Nijmegen, The Netherlands}
\affiliation{Theoretical Physics and Applied Mathematics Department, Ural Federal University, 620002 Yekaterinburg, Russia}
\author{Mikhail Danilov}
\affiliation{Institute of Theoretical Physics, University of Hamburg, 20355 Hamburg, Germany}
\author{Sergey Brener}
\affiliation{Institute of Theoretical Physics, University of Hamburg, 20355 Hamburg, Germany}
\affiliation{The Hamburg Centre for Ultrafast Imaging, Luruper Chaussee 149, 22761 
Hamburg, Germany}
\author{Malte~Harland}
\affiliation{Institute of Theoretical Physics, University of Hamburg, 20355 Hamburg, Germany}
\author{Alexander I. Lichtenstein}
\affiliation{Institute of Theoretical Physics, University of Hamburg, 20355 Hamburg, Germany}
\affiliation{The Hamburg Centre for Ultrafast Imaging, Luruper Chaussee 149, 22761 
Hamburg, Germany}
\affiliation{Theoretical Physics and Applied Mathematics Department, Ural Federal University, 620002 Yekaterinburg, Russia}
\author{Mikhail I. Katsnelson}
\affiliation{Radboud University, Institute for Molecules and Materials, 6525AJ Nijmegen, The Netherlands}
\affiliation{Theoretical Physics and Applied Mathematics Department, Ural Federal University, 620002 Yekaterinburg, Russia}
  \blfootnote{A.A.B. and M.D. contributed equally to this work}

\begin{abstract}
A considerable success in phenomenological description of high-T$_{\rm c}$ superconductors has been achieved within the paradigm of Quantum Critical Point (QCP) - a parental state of a variety of exotic phases that is characterized by dense entanglement and absence of well-defined quasiparticles.
However, the microscopic origin of the critical regime in real materials remains an open question.
On the other hand, there is a popular view that a single-band $t-t'$ Hubbard model is the minimal model to catch the main relevant physics of superconducting compounds. Here, we suggest that emergence of the QCP is tightly connected with entanglement in real space and identify its location on the phase diagram of the hole-doped $t-t'$ Hubbard model. To detect the QCP we study a weighted graph of inter-site quantum mutual information within a four-by-four plaquette that is solved by exact diagonalization. We demonstrate that some quantitative characteristics of such a graph, viewed as a complex network, exhibit peculiar behavior around a certain submanifold in the parametric space of the model. 
This method allows us to overcome difficulties caused by finite size effects and to identify the transition point even on a small lattice, where long-range asymptotics of correlation functions cannot be accessed.

\end{abstract}

\maketitle
\subsection*{Introduction}
The phenomenon of high-temperature superconductivity (HTSC) still remains very puzzling after more than thirty years since the discovery of superconducting copper-oxide compounds  \cite{Bednorz}. Serious hopes for the understanding of this phenomenon are related to the concept of a quantum critical point (QCP)  \cite{Sachdev_book, Sachdev_new}, - an exotic state of matter that exhibits scale invariance and lacks long-lived quasiparticles, and thus cannot be described by means of conventional Fermi-liquid theory. Contemporary discussions of observed properties of HTSC are frequently organized around this concept \cite{Hussey, Michon}. Precise nature of this critical point is still unclear, -- different studies relate it to charge density waves \cite{Arpaia}, nematic  \cite{Auvray}, or antiferromagnetic fluctuations  \cite{Wang}. However, its theoretical treatment can be conducted universally, though it requires a change of basic mathematical tools: the diagrammatic approach, the main apparatus of quantum many-body theory during the last sixty years \cite{AGD,Mahan}, is not really fitted to the description of systems lacking quasiparticles.
A paradigmatic shift in studying strongly coupled systems near the QCP has occurred when it was realized that the anti de Sitter/Conformal field theory (holographic) correspondence \cite{Maldacena} can be used to analyze certain universal phenomenological properties of correlated electronic matter in the regime where the traditional Fermi-liquid picture breaks down \cite{Zaanen,Hartnoll_book}. While the number of direct experimental evidences of quantum critical points in high-Tc superconducting materials is limited, assuming its existence and employing the methods of holography allowed to resolve within a relatively short time frame a number of puzzles that remained perplexing for decades \cite{Anderson}. The correspondence provided an explanation for the linear-$T$ scaling of DC resistivity in the normal state of cuprates \cite{Legros} (known as strange metals), relating it to general hydrodynamic properties of systems with minimal viscosity proportional to the thermodynamic entropy \cite{Davison}. It was shown \cite{Blake} that the Hall angle, - the temperature dependent ratio of the Hall and DC conductivities, $\tanh{\theta_H}=\sigma_{xy}/\sigma_{xx}\sim 1/T^2$, can be naturally interpreted in terms of a two-constituent quantum liquid, where the regular quasiparticles and the critical sectors give independent contributions to the conductivity, leading to an anti-Matthiessen rule for transport. A new mechanism of the interaction-driven metal-insulator transition that causes anisotropic localization has been suggested \cite{Hartnoll}, and it appears to be fully in line with the localization of conducting electron gas in two-dimensional CuO planes, while the conductivity in the orthogonal direction is suppressed. Other phenomena, such as the formation of Fermi arcs seen in the angle-resolved photoemission spectra of high-Tc compounds \cite{Iliasov}, or charge density waves \cite{Krikun} also fit pretty naturally into the context of quantum criticality.

The main problem of this approach is its purely phenomenological character. It cannot explain by itself why the high-Tc compounds, contrary to the most of interesting condensed matter systems, do not behave as the Fermi liquid but instead are characterized by minimal quantum viscosity and other fancy properties. Such an explanation requires an analysis of electronic structure of specific materials.

Current understanding of high-T$_{\rm c}$ superconductivity in cuprates assumes a crucial role of strong electron correlations \cite{Anderson,Dagotto,Millis,Scalapino}, which are taken into account within a particular minimal model that was formulated \cite{Andersen1} on the basis of the density functional band structure of cuprates, - the single-band $t-t'$ Hubbard model on a square lattice given by the Hamiltonian
\begin{equation}
\label{hubbard}
    H=-t\sum\limits_{\langle i,j\rangle, \sigma}c^\dagger_{i,\sigma} c_{j,\sigma}-t'\sum\limits_{\langle\langle l,k\rangle\rangle, \sigma}c^\dagger_{l,\sigma} c_{k,\sigma}+h.c. +U\sum\limits_{i} n_{i,\uparrow}n_{i,\downarrow},
\end{equation}
where, the first sum is taken over the pairs $\langle i,j \rangle$ of nearest neighbors, the second one - over the pairs $\langle \langle l,k \rangle\rangle$ of next-to-nearest (diagonal) neighbors, $c_{i,\sigma}$ is the electron annihilation operator, and the on-site occupation operator is $n_{i,\sigma} = c^\dagger_{i,\sigma}c_{i,\sigma}$. For convenience, in what follows we express all energies in units of $t$ .

In an attempt to connect the phenomenological and the microscopic levels of description of HTSC, we shall focus on the $t-t'$ Fermi-Hubbard model and try to detect QCP on its phase diagram.

Correlation effects beyond the band structure approximation in this model have been thoroughly analyzed with different methods  \cite{Maier, Haule, Khatami, Gull, Civelli}, and there are a number of good indications that it captures all the relevant features of cuprate superconductors. In a series of papers \cite{IKK_parquet,Metzner1,Metzner2,Metzner3,Metzner4,Honerkamp}, perturbative renormalization group studies of the model have been conducted, and the emergence of the superconducting order parameter and the competition between superconductivity and antiferromagnetism were demonstrated. In particular \cite{Metzner4}, it was argued that the next-to-nearest neighbor hopping $t'$ plays a crucial role in the stabilization of superconductivity.
A complementary approach is based on the cluster dynamical mean-field studies
which consider a 2-by-2 plaquette as an elementary unit \cite{LK2000}.
Recently \cite{Harland1}, it was noticed that this plaquette has a very special electronic structure for the parameters and the electron occupation number typical for the the optimal doping regime in YBa$_2$Cu$_3$O$_7$ ($t'=-0.3$, $U\simeq6$), with an ``accidental'' degeneracy of many-electron energy levels and formation of the soft fermion mode due to this degeneracy.  The pseudogap forms via this mode by a mechanism of the Fano antiresonance, and the superconducting d-wave susceptibility dominates over other instability channels. This behavior was interpreted in terms of formation of a local plaquette valence bond state. On a larger scale, the ground state of the model has been analyzed by means of density matrix renormalization group \cite{Devereaux1} (DMRG) (see also \cite{Kivelson} for the related studies of its cousin, $t-J$-model), and additional arguments in favor of stabilization of superconductivity by the next-to-nearest neighbor hopping were provided. In turn, at temperatures above the superconducting phase transition, determinantal Monte Carlo computations \cite{Devereaux2} demonstrated that the DC resistivity exceeds the Mott-Ioffe-Regel limit and scales linearly with temperature.

The search for the QCP in the $t-t'$ Hubbard model has been performed within the dynamical cluster approximation \cite{Khatami}, and its existence has been proven by studying thermodynamics properties of the model at finite temperature and their further extrapolation to $T=0$. However, it is tempting to get a deeper insight into the microscopics of the QCP and demonstrate its emergence due to interactions of electrons at low temperatures.

Since large scale simulations of the fermionic Hubbard model away from half-filling are challenging because of the sign problem, it is natural to ask whether we can extract any information about the tendency to form critical states out of small cluster solutions obtained by means of exact diagonalization. At first, this goal does not seem realistic since studying systems in the critical regime unavoidably requires dealing with long-range correlations, while all the microscopic precursors of the transition on small lattices would be washed out by the finite-size effects. However, it is useful to bear in mind that, in the context of many-body quantum dynamics, the concept of entanglement and the phenomenon of collective emergence go hand in hand. An archetypical example of such relation is the Cooper pairs in the BCS theory of superconductivity: while the ground state wavefunction has a form of a product state of the Cooper pairs, each pair itself is a two-body entangled system. Therefore it is natural to expect that major transitions in phenomenological properties of many-body systems would be reflected in the patterns of entanglement, and quantum criticality should leave its fingerprint on all scales, not only in the deep infrared limit. A nice example of how fruitful this way of thinking can be was given in Refs. \cite{Tremblay1, Tremblay2}, where entanglement measures were used to determine universality class of the Mott transition in the 2d Hubbard model.

Recently, a novel approach to phase transitions in quantum lattice models based on complex network theory has been suggested \cite{Carr1, Carr2}. It was noticed that a particular structure that can be computed with relative ease and appears to be very sensitive to reconfigurations of the quantum state is the network of quantum mutual information.
\begin{figure}[t!]
\begin{center}
\includegraphics[width=0.5\linewidth]{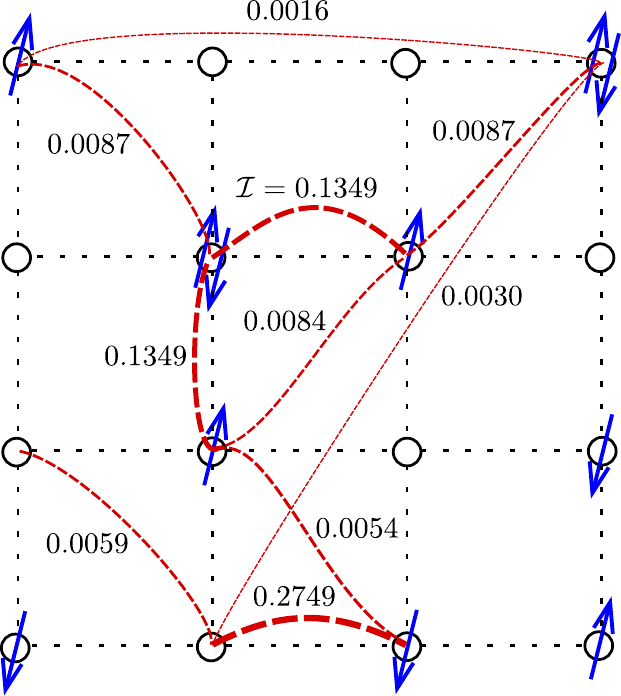}
\end{center}
\caption{An artistic view of the mutual information complex network defined on the Hubbard lattice. While the network is fully connected, for illustrative purposes, only some of the network links are shown. The shown values of inter-site mutual information correspond to the case of non-periodic boundary conditions, $(6,6)$ sector, $U=7.5$, $|t'|=0.3$.\label{fig:network}
}
\end{figure}
The mutual information between two subsystems $A$ and $B$ of a larger systems is defined as
\begin{equation}
\label{entropy}
    {\cal I}_{AB}=S_A+S_B-S_{A\cup B},
\end{equation}
where $S_A=-\Tr{\rho_{A}\log\rho_{A}}$ is the von Neumann entropy, and $\rho_A=\Tr_{\bar{A}}\rho$ is the density matrix of subsystem $A$. Then we can associate a weighted graph with a state of a quantum lattice system, e.g. the Hubbard model, by considering the lattice sites $i=1\dots N$, where $N$ is the number of sites, as nodes of the graph, and the values of pairwise inter-site mutual information $I_{i j}$ play the role of weights on the graph links (see Fig. 1). This representation is appealing for the following reason. Once a wave function on the lattice is known, it is easy to compute the entanglement entropy of a pair of sites and thus the mutual information. At the same time, such a network contains information of quantum correlations which could be very important to understand the dynamics of strongly correlated systems.
In the cases of the transverse field Ising and the Bose-Hubbard models in 1d, it was demonstrated that certain characteristics of the mutual information network can be used to detect quantum phase transitions \cite{Carr1, Carr2}. Namely, behavior of the following functions upon changing parameters of the models has been studied:
\begin{itemize}
    \item {\bf Clustering} of a weighted graph is defined as
        \begin{equation}
        C=\frac{\Tr {\cal I}^3}{\sum_{j\neq i}^N \sum_{i=1}^N \left[{\cal I}^2 \right]_{ij}},
    \end{equation}
    where $N$ is the total number of sites in the lattice, and $\cal I$ is the $N \times N$ matrix of inter-site mutual information. One can see that this quantity maximizes on graphs with a lot of three-link loops with high weights. For the cases studied in Ref. \cite{Carr1}, it was shown that it serves as sensitive detector that exhibits a clear dip at the phase transition point. A natural explanation of this fact is that, at the criticality, one can expect the corresponding network to be scale-free, and for generic scale-free networks clustering is usually quite low  \cite{scalefreeclustering}.
    \item {\bf Disparity} of a single node in a network is defined as a measure to capture how non-uniformly weights on the links attached to this node are distributed:
    \begin{equation}
        Y_{i}=\frac{\sum_{j=1}^N \left({\cal I}_{ij}\right)^2}{\left(\sum_{j=1}^N {\cal I}_{ij} \right)^2}
    \end{equation}
    For example, if the node has the same value of mutual information with all the other nodes of the network, its disparity would be $Y_i=1/(N-1)$, while if it correlates only with one neighbor, the disparity maximizes as $Y_i=1$. Physically speaking, high disparity of a lattice site means that it tends to correlate only with a few other sites, and ``factorize out'' of the rest of the system. In the context of quantum many-body physics such a behavior would be typical for states that can be nearly decomposed into product states. On the other hand, low disparity means that the site correlates with a large number of degrees of freedom.
    \item {\bf Density} is an overall characteristic of a network given by
    \begin{equation}
        D=\frac{1}{N\left(N-1\right)} \sum\limits_{i,j=1}^N {\cal I}_{ij},
    \end{equation}
    i.e. it is the averaged fraction of all the weights (mutual information values) of the network. To gain more intuition on what properties of the many-body quantum state it reflects, we shall estimate an upper bound on this measure. If site $i$ of the network is maximally entangled with the rest of the system, its entanglement entropy equals $S_i=\ln d = \ln 4$, where $d=4$ is dimension of the local on-site Hilbert space in Hubbard model. On the other hand, mutual information monogamy theorem implies that $2S_i\geq \sum\limits_{i,j} {\cal I}_{ij}$  \cite{monogamy}, leading to
    \begin{equation}
        D \leq \frac{2}{N\left(N-1\right)} \sum\limits_{i=1}^N S_i \leq \frac{2\ln 4}{N-1}\xrightarrow[N\rightarrow \infty]{} 0 \label{eq:Dupper}
    \end{equation}
    i.e. the mutual information network is generally sparse even if the system is highly entangled. Note that bound \eqref{eq:Dupper} can be saturated in physically very distinct cases. $D$ is maximal if either each single site is maximally entangled with just one partner site, and the state as a whole decomposes into a product of Bell pairs, or if the entanglement between the site and the rest of the system is homogeneously scrambled over all the sites. To distinguish between such configurations one has to refer to the disparity which we defined above.
    \item {\bf Pearson correlations} measure how much two nodes $i$ and $j$ of a network differ from each other:
    \begin{gather}
        r_{ij} = \frac{\sum_{k=1}^N\left({\cal I}_{ik}-\langle{\cal I}_i\rangle \right)\left({\cal I}_{jk}-\langle{\cal I}_j\rangle\right)}{\sqrt{\sum_{k=1}^N\left({\cal I}_{ik}-\langle{\cal I}_i\rangle \right)^2}\sqrt{\sum_{k=1}^N\left({\cal I}_{ik}-\langle{\cal I}_i\rangle \right)^2}}, \\
        \langle {\cal I}_i \rangle = \frac{1}{N}\sum_{j=1}^N {\cal I}_{ij} \nonumber
    \end{gather}
    In Ref. \cite{Carr1} Pearson correlations of neighboring nodes were shown to develop a cusp around the phase transition point.
\end{itemize}
For one-dimensional Ising and Bose-Hubbard models \cite{Carr1}, this approach to detecting quantum phase transitions points was successfully applied for systems of $\sim 10^2$ sites, and was demonstrated to be very robust upon finite-size effects. In the two-dimensional case, we are limited by much smaller system sizes (we perform exact diagonalization for a 4-by-4 plaquette), and should not expect our results to be free from finite-size artifacts. Still, as we shall see in the next section, the network measures exhibit clearly distinguishable features on a submanifold of the $t-t'$ Hubbard model parametric space. In particular, this submanifold includes the level-crossing point observed in a 2-by-2 plaquette for the choice of parameters corresponding to YBa$_2$Cu$_3$O$_7$ superconductor \cite{Harland1}.

\subsection*{Results}
\begin{figure}[h!]
\centering
\includegraphics[width=0.69\linewidth]{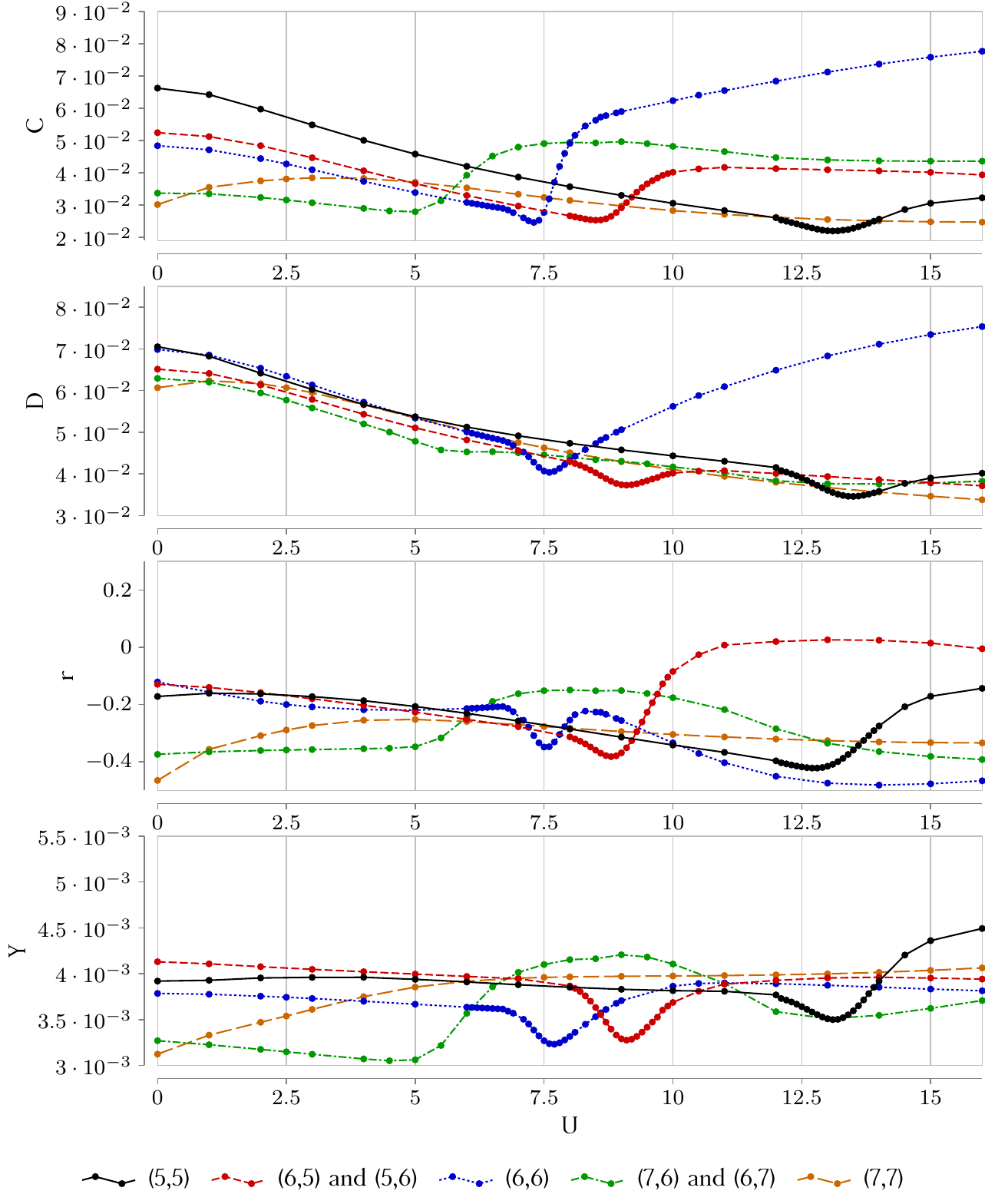}
\caption{Characteristics of the mutual information complex network, -- clustering $C$, density $D$, Pearson correlation $r$ between neighboring sites in the middle of the 4-by-4 plaquette, and disparity $Y$ of a site in the middle of the plaquette, -- as functions of the on-site Coulomb repulsion $U$ computed in different sectors for non-periodic boundary conditions. The hopping is $t'=-0.3$, the inverse temperature is $\beta = 100$. \label{fig:complex_network_lowT}
}
\end{figure}
We have computed the complex network measures discussed above across the space of parameters of a 4-by-4 $t-t'$ Hubbard plaquette.
Within each fixed particle number sector, from $(5,5)$ ($37,5\%$ hole doping) to $(7,7)$ ($12,5\%$ hole doping), we scan over $t'$ and $U$, Fig. \ref{fig:complex_network_lowT}. As an indicative value, we take $t'=-0.3$, which is estimated to be the next-neighbor hopping in the Hubbard model of YBaCuO compounds, and search for transition points around it. The temperature is fixed to $1/T=\beta=100$ (all energies are expressed in the units of $|t|$), and the system is studied in the canonical ensemble.

\begin{figure}[t!]
\centering
\includegraphics[width=0.49\linewidth]{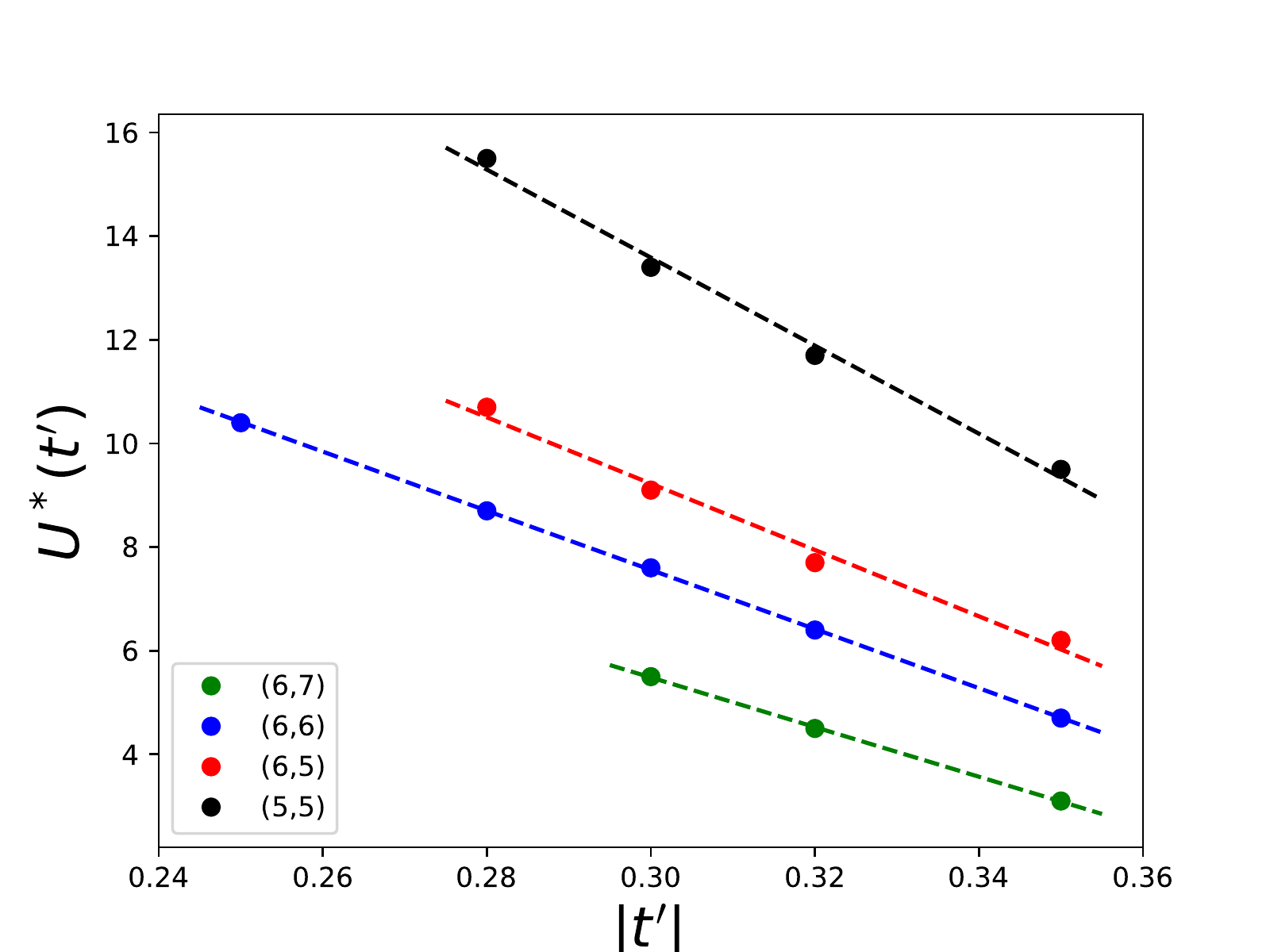}
\caption{Dependence of the critical Coulomb repulsion $U^*$ on the next-neighbor hopping $t'$, as the latter is varied in the range $t' \in \left[-0.35,\,-0.25 \right]$ for non-periodic boundary conditions at inverse temperature $\beta = 100$. The points correspond to locations of disparity minimum.  \label{fig:linear_law}
}
\end{figure}

We assume that a transition point is evident if all the measures exhibit some clear features around the same point. Accepting this criterion, we can claim with a high confidence that, for non-periodic boundary-conditions, there is a family of transition points in each sector (except $(7, 7)$) forming a nearly perfect straight line in the $t'-U$ plane that extends in a certain range of $t'$ \footnote{For a more detailed picture of how complexity measures behave at different values of $t'$, see Ref. \cite{Supplemental}}, Fig. \ref{fig:linear_law} (for too small $|t'|$ the signs of criticality are faded away from the complexity measures). Moreover, for different values of hole doping, all these lines have very similar slope. This can be interpreted as that phase transition occurs on a 2d manifold in the 3d parametric ($U/t$, $t'/t$, particle number) space of the model. 

We stick to non-periodic boundary conditions for the reason that the mutual information network has a richer structure in that case. If periodic boundary conditions are imposed, all lattice sites are identical, and every site has only five inequivalent connections to others, making the  ${\cal I}_{ij}$ matrix highly degenerate. Hence, the corresponding network structures are constrained by symmetries and much less sensitive to variations of the model parameters. 
Still, we would like to note that, when boundary conditions are changed for periodic ones, all the phase transition lines are smeared out with the only exception of the $(6,6)$ sector which corresponds to the hole doping of $\delta=25\%$. For the latter, only the concrete values of Coulomb repulsion $U$ gets shifted \cite{Supplemental}. While we do not expect the information network constructed for periodic boundary to be sensitive enough to properly detect phase transitions, it is interesting to note that this single sector where the transition is evident for both choices of b.c. is the same as the one where level-crossing associated with formation of the pseudogap via Fano antiresonance occurs in a 2-by-2 plaquette \cite{Harland1}.

\begin{figure}[t!]
\centering
\includegraphics[width=0.79\linewidth]{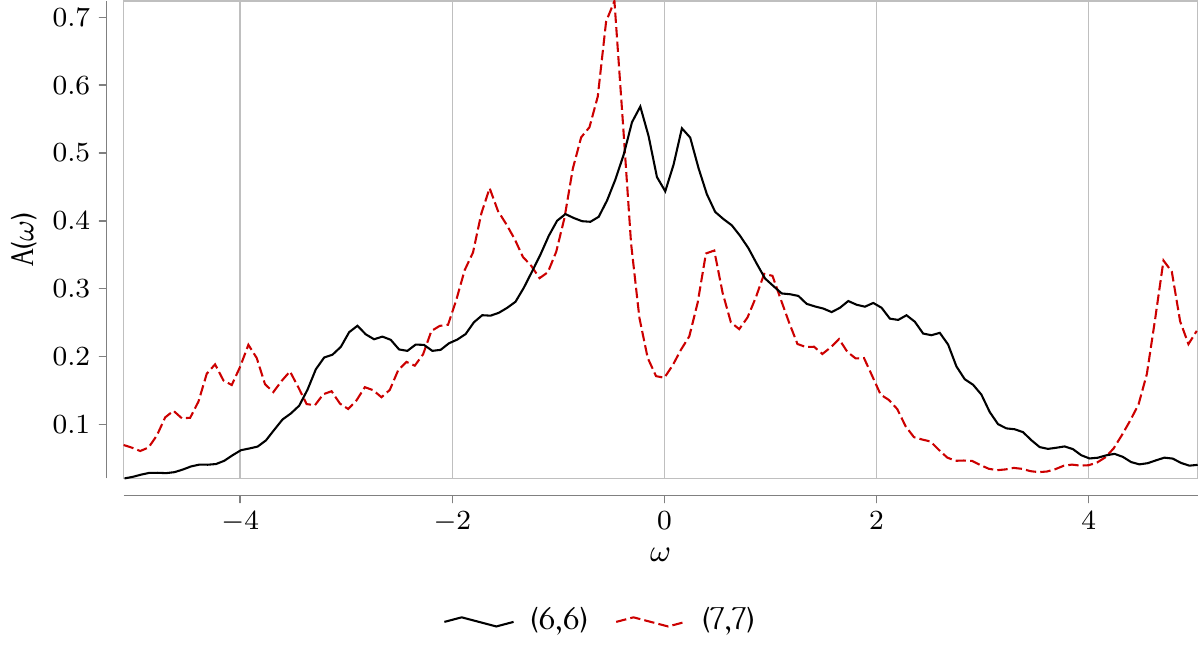}
\caption{The spectral density of states $A(\omega)$ defined as the imaginary part of the retarded Green function, computed for $U=7.5$ with non-periodic boundary conditions ($\omega=0$ corresponds to the Fermi energy). One can see the pseudogap formation around $\omega=0$ near the quantum critical point. Its interpretation in terms of the Fano antiresonance due to formation of a ``soft fermion'' mode was given in Ref. \cite{Harland1}} \label{fig:dos}
\end{figure}

At the same time, in the density of states (d.o.s.) the transition point is (almost) invisible. Some minor peculiarity at the quantum critical point is visible in the density of states at $t'=-0.3$. Around the transition point identified by means of the complex network theory ($U=7.5$, sector (6,6)) the $\omega=0$ peak in the d.o.s. starts splitting and the pseudogap emerges, see Fig. \ref{fig:dos}. Further decrease of the hole doping leads to enhancement of the gap. The particular role of $U$ in this transition is less clear, as the d.o.s. profile varies very mildly upon changing $U$,
%The only peculiarity one can spot is that the emerged peaks become symmetric when passing the $U\simeq 7.5$ point in the (6,6) sector. 
and it would be safer to claim that the spectral properties are not sensitive to the discussed quantum phase transition. Ideologically, this situation is somewhat similar to the Anderson localization in disordered systems which is a clear example of a phenomenon that cannot be detected on the level of the average Green's functions  \cite{Disordered}.
\subsection*{Discussion.}
By associating the quantum state of the $t-t'$ Hubbard model with a weighted network of inter-site mutual information, for different values of the next-neighbor hopping $t'$, we have found a set of transition lines in the $U-t'$ plane of the model parametric space, where characteristics of the network have a clearly distinguishable cusp. Such a behavior was previously shown to be an indication of quantum phase transitions in different one-dimensional models  \cite{Carr1, Carr2}.
The modern experimental understanding of the putative QCP in cuprates tells that it must be associated with the emergence of the pseudogap phase  \cite{Hussey}. For example, for YBaCuO compounds the onset of pseudogap was experimentally demonstrated to happen at hole doping $\delta\simeq 22\%$  \cite{Sato}. The hole doping $\delta=25\%$ is the closest value one can get for a 4-by-4 cluster (the $(6,6)$ sector), and, interestingly, it is exactly the sector where the complex network measures demonstrate the most robust transition features. The particular values of the on-site Coulomb repulsion is affected by the finite size effects, and estimated to be about $U\simeq 7\,-\,8$.
 At the same time, no peculiarity is seen in the density of states at the transition point, apart from a slight splitting of d.o.s. around the Fermi level,
which might be a good indication that the low-order correlation functions that define the spectral and the response properties of the system could be blind to restructuring of many-body quantum states, and does not contain enough information on the role of quantum correlations behind phase transitions in electron systems.

Both the strength and the weakness of the employed approach is that it helps to identify {\it any} critical point while being ignorant about its nature. Therefore we can claim the existence of a manifold of QCP in the $t-t'$ Hubbard model with a high confidence, but we cannot deduce what order parameter of the corresponding transition is. Still, we tend to relate the observed transition to the critical point discussed in  \cite{Harland1}, where it was associated with emergence of soft fermion modes.

Within the exact diagonalization approach, we were able to consider only a small cluster, where one unavoidably has to deal with strong finite size effects. It would be interesting to conduct a similar analysis for a larger system. The state-of-the-art DMRG allows to study quasi-2d stripes as large 4-by-64 atomic sites \cite{Devereaux1, Kivelson, Devereaux2}, and we hope to apply the complex network approach to such systems in the future.

\subsection*{Methods.}
In this section we give the relevant technical details of the calculation of the entanglement measures defined above. The first step is to diagonalize the Hubbard model~(\ref{hubbard}) for a 4-by-4 cluster. This can be done either for a periodic or a non-periodic model. The diagonalization is performed using the Lanczos algorithm with 200 Krylov basis vectors  \cite{Lanczos}. The particle number and spin conservation laws are used so that the diagonalization can be restricted to a sector with a fixed number of up- and down-spins. Those eigenstates with the corresponding eigenvectors are then used to calculate the reduced density matrices for each possible pair of sites as well as for each single site.

The reduced density matrix is computed using its definition that can be symbolically written as:
\begin{equation}
    \rho_A({a,a'})=\frac{1}{Z}\sum_n e^{-\beta E_n}\Tr_{\bar{A}}\left|\psi_{n,(a,\bar{a})}\right\rangle\left\langle\psi_{n,(a',\bar{a})}\right|.
\end{equation}

Here $a,a'$ denote the many-particle (Fock) basis states describing the subsystem $A$ we calculate the density matrix for, $\bar{a}$ stands for the many-particle basis state of the complementary subsystem $\bar{A}$, thus a couple of those $(a,\bar{a})$ denotes a basis Fock state for the whole cluster explicitly split into two parts. As before, $n$ stands for a particular eigenvector, the density matrices for given eigenstates are weighted with the Boltzmann factors corresponding to their energies. In a given sector for a given set of parameters we use the Boltzmann factor cut-off of 1\% meaning $e^{(E_0-E_{k})\beta}>10^{-2}$, where $E_0$ is the ground state energy and $E_{k}$ is the energy of the highest ($k$th) level taken into account. Note that while performing the partial trace over $\bar{A}$ one has to correctly account for the fermionic commutation relations. To this aim one has to effectively change the numeration of sites so that the sites for which we calculate the density matrix stand first. Explicitly it means that each component of an eigenvector, corresponding to a given basis state of the cluster, gets a factor determined as the parity acquired while "dragging" the occupied sites of $A$ to the beginning past the occupied states of $\bar{A}$. In other words for each basis vector one takes each occupied site from $A$ and for each occupied spin component counts the number of same spin occupied sites from $\bar{A}$ standing before the considered site in the original numeration. Summing up the parities of those numbers for all occupied sites and spins from $A$ one gets the parity that is assigned to a given basis vector with respect to the subsystem $A$. Having multiplied the eigenvector components with the acquired parities one finally performs the partial trace over the complementary subset $\bar{A}$.

Given the reduced density matrix, we first calculate the von Neumann entropy of a given subsystem and then, with~Eq.~(\ref{entropy}), the mutual information for each pair of sites, that serves as the basis for our network.

The $\omega$-dependent Green function is given by:
\begin{equation}
    G_{i,\sigma}(\omega)=\frac{1}{Z}\sum_{m,n}\frac{|\left\langle m\right|c^\dagger_{i,\sigma}\left|n\right\rangle|^2}{\omega+E_n-E_m}\left(e^{-\beta E_n}+e^{-\beta E_m}\right).
\end{equation}
Here $m,n$ denote the eigenstates of the system, $i$ and $\sigma$ denote a given site and spin (in the paramagnetic case the answer is spin-independent), $E_n$ is the energy of the $n$-th state, and $Z=\sum_m e^{-\beta E_m}$ is the partition function. Note that $m$ and $n$ necessarily belong to different sectors.

The Green function is then used to calculate the spectral density of states, which is defined as
\begin{equation}
 A(\omega) = - \frac{1}{\pi}\mbox{Im} \Tr_{i,\sigma} G_{i,\sigma}(\omega).
\end{equation}
To perform numerical computations, delta-peaks in the Green function are broadened with $\delta=\pi/\beta$.
\vskip10pt
{\it Acknowledgements --} the authors thank Lincoln Carr for inspiring discussions.

{\it Competing Interests --} the Authors declare no Competing Financial or Non-Financial Interests.

{\it Author contributions --} A.A.B., M.I.K. and A.I.L. designed the project and directed it with the help of S.B., M.D. and M.H. performed the calculations. A.A.B., M.I.K. and S.B. wrote the manuscript. All authors contributed to discussions.

{\it Funding --} the work of A.A.B. was supported by Russian Science Foundation, Grant No. 18-12-00185. M.I.K. and A.I.L. acknowledge a support by European Research Council via
Synergy Grant 854843 - FASTCORR. M.H., S.B. and A.I.L. acknowledge support by the Cluster of Excellence 'Advanced Imaging of Matter' of the Deutsche Forschungsgemeinschaft 
(DFG) - EXC 2056 - project ID 390715994.
 
{\it Data Availability --} the data that support the findings of this study are available from the corresponding author upon reasonable request.

\newpage
\begin{center}
{ \bf Supplemental Material}
\end{center}

\setcounter{figure}{0}
\makeatletter
\renewcommand{\thefigure}{S\@arabic\c@figure}
\makeatother

\vskip1cm

In this Supplemental Material, we provide results for the dependence of complex network measures on Coulomb repulsion $U$ at different values of next-neighbor hopping $t'$. We also show figures for the periodic boundary conditions for the $(6,6)$ sector, - the only one where the transition point is evident for periodic b.c.
\vskip 10pt
\begin{figure}[h!]
\centering
\includegraphics[width=0.49\linewidth]{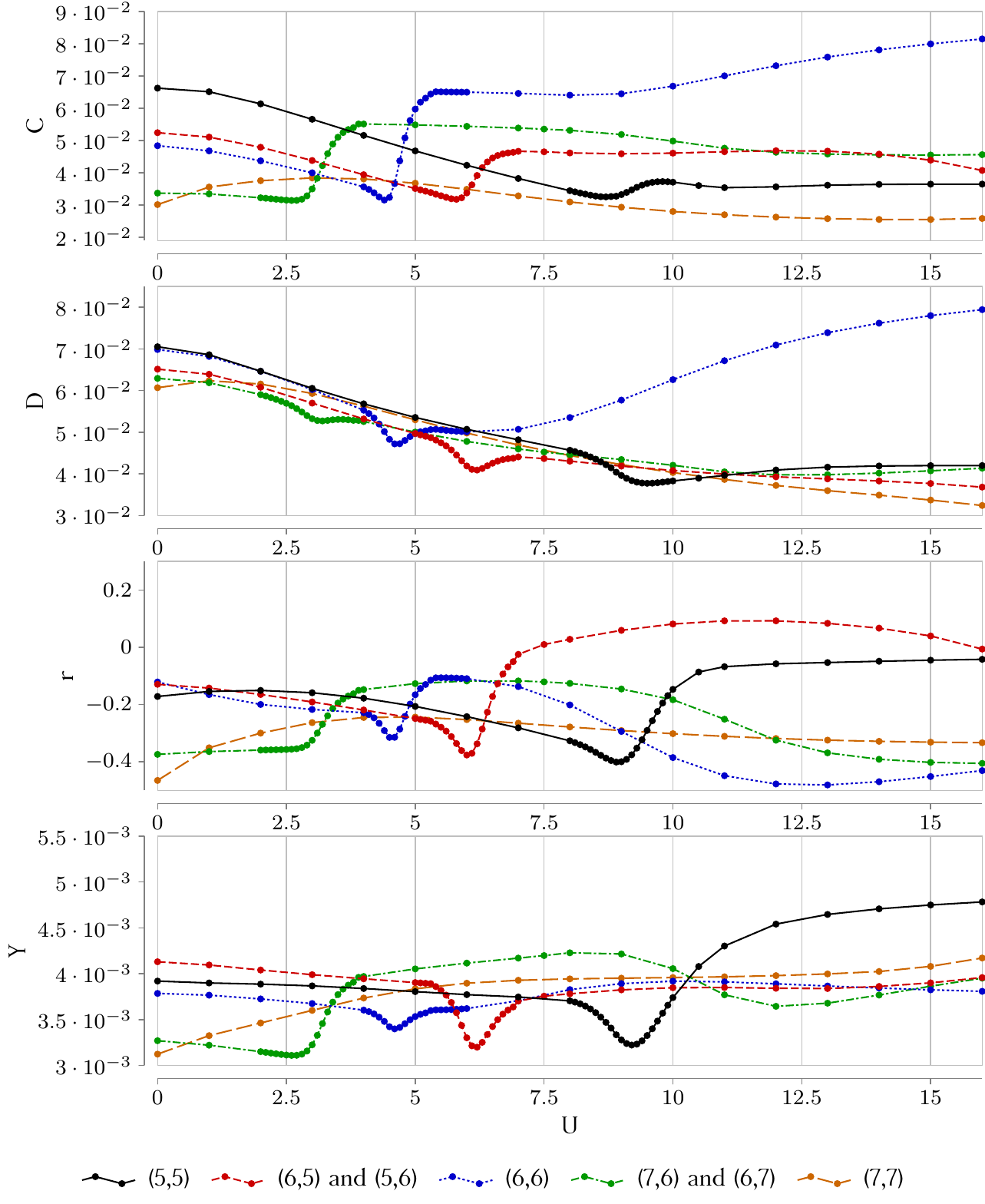}
\includegraphics[width=0.49\linewidth]{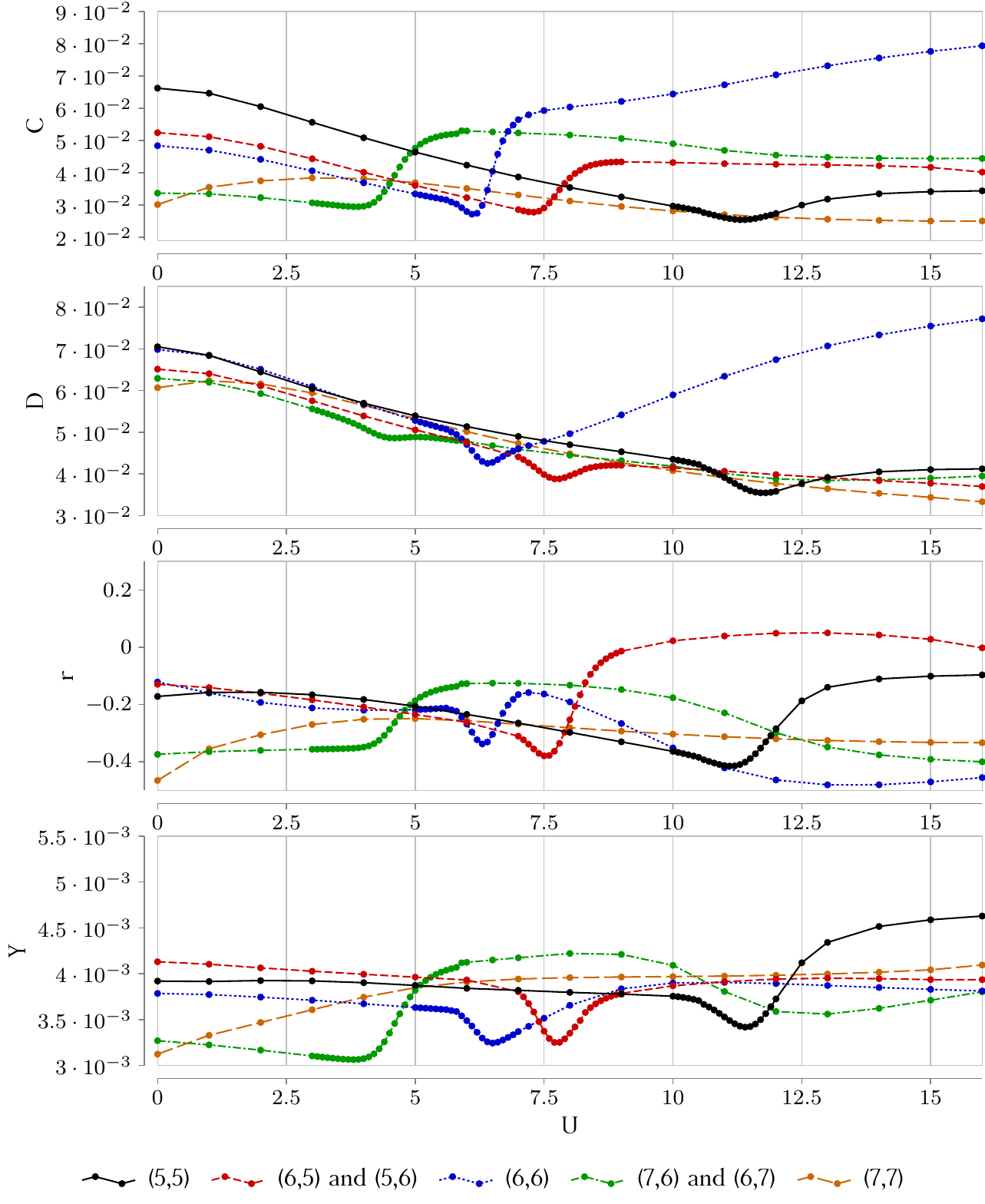}
\caption{Characteristics of the mutual information complex network, -- clustering $C$, density $D$, Pearson correlation $r$ between neighboring sites in the middle of the 4-by-4 plaquette, and disparity $Y$ of a site in the middle of the plaquette, -- as functions of the on-site Coulomb repulsion $U$ computed in different sectors for non-periodic boundary conditions. The hopping at $t'=-0.35$ (left) and $t'=-0.32$ (right), the inverse temperature is $\beta = 100$.  \label{fig:complex-1} 
}
\end{figure}

\begin{figure}[h!]
\centering
\includegraphics[width=0.49\linewidth]{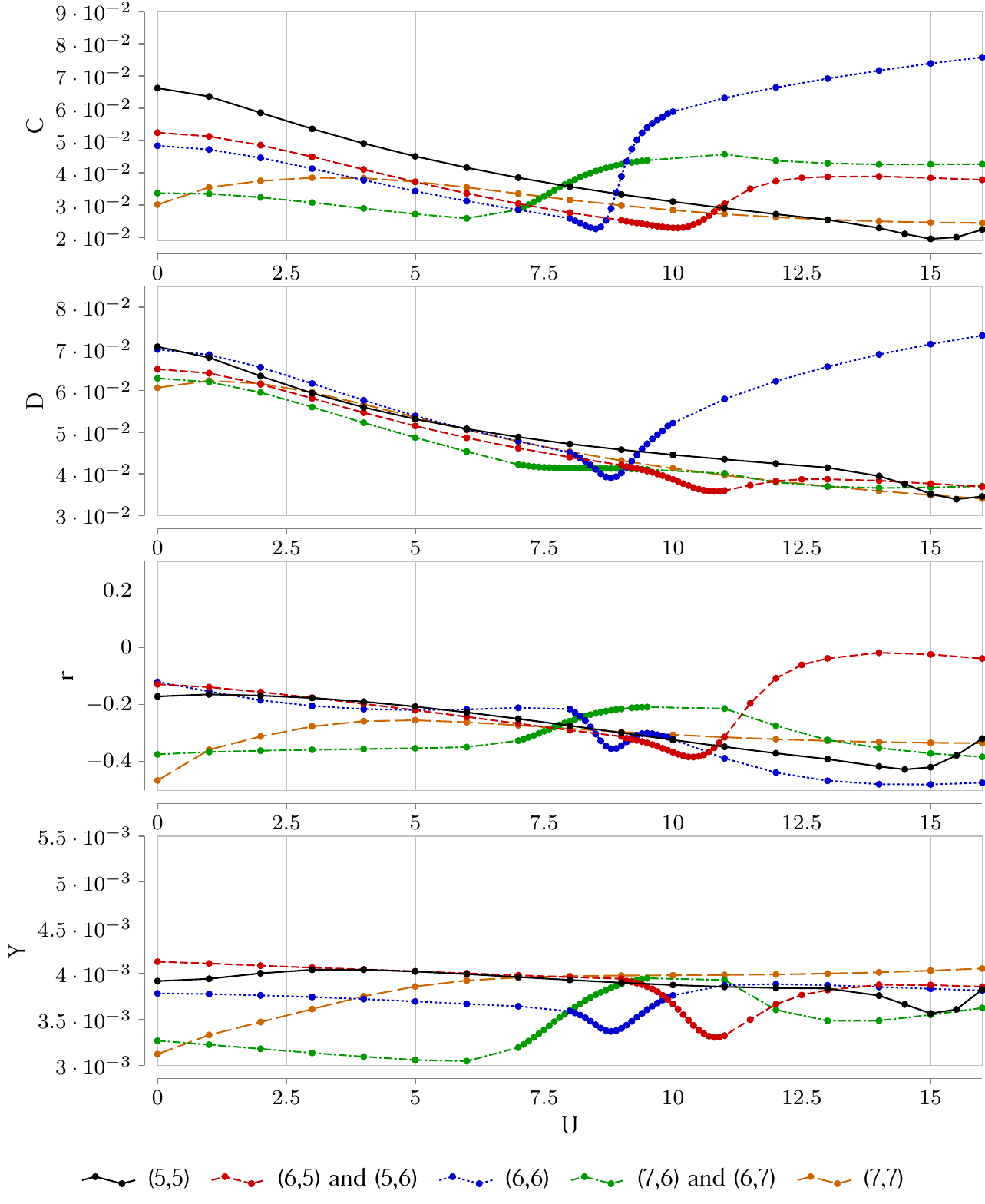}
\includegraphics[width=0.49\linewidth]{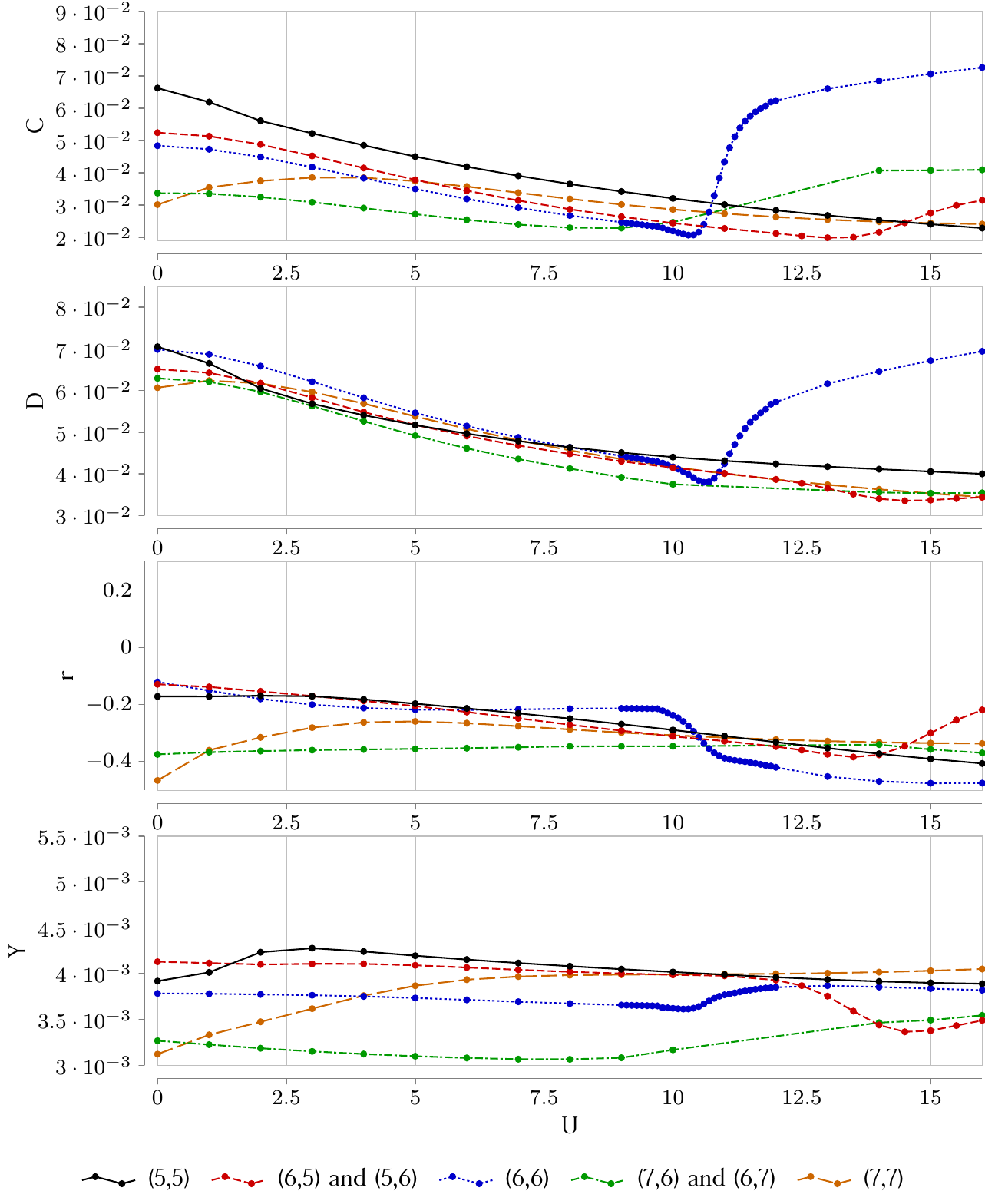}
\caption{Characteristics of the mutual information complex network, -- clustering $C$, density $D$, Pearson correlation $r$ between neighboring sites in the middle of the 4-by-4 plaquette, and disparity $Y$ of a site in the middle of the plaquette, -- as functions of the on-site Coulomb repulsion $U$ computed in different sectors for non-periodic boundary conditions. The hopping at $t'=-0.28$ (left) and $t'=-0.25$ (right), the inverse temperature is $\beta = 100$.  \label{fig:complex-2} 
}
\end{figure}

\begin{figure}[h!]
\centering
\includegraphics[width=0.69\linewidth]{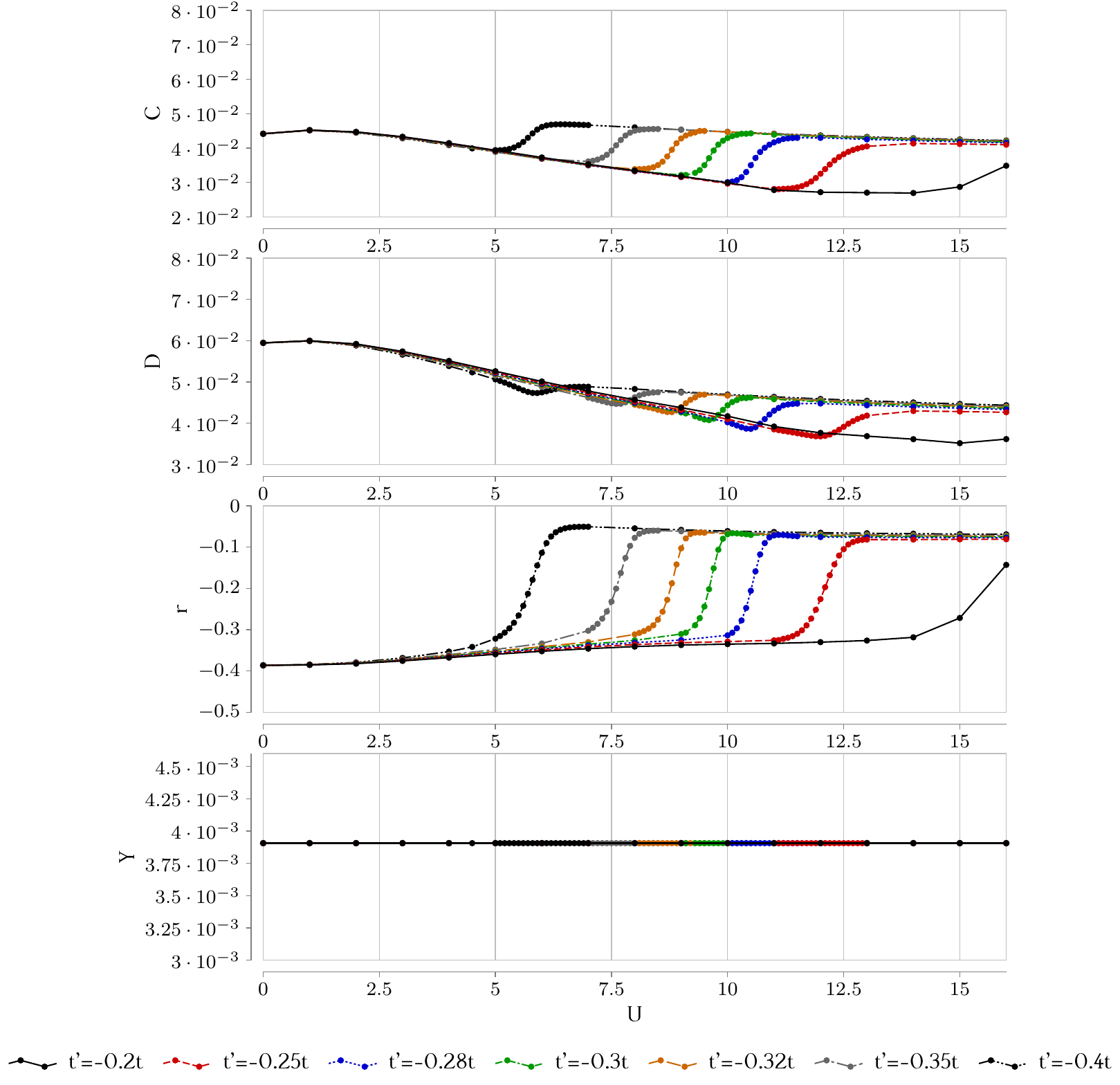}
\caption{Characteristics of the mutual information complex network, -- clustering $C$, density $D$, Pearson correlation $r$ between neighboring sites in the middle of the 4-by-4 plaquette, and disparity $Y$ of a site in the middle of the plaquette, -- as functions of the on-site Coulomb repulsion $U$ computed in the (6,6) sector for periodic boundary conditions. It is the only sector where mild features of the phase transition survive upon changing the boundary conditions.  \label{fig:complex-2} 
}
\end{figure}

\begin{figure}[h!]
\centering
\includegraphics[width=0.49\linewidth]{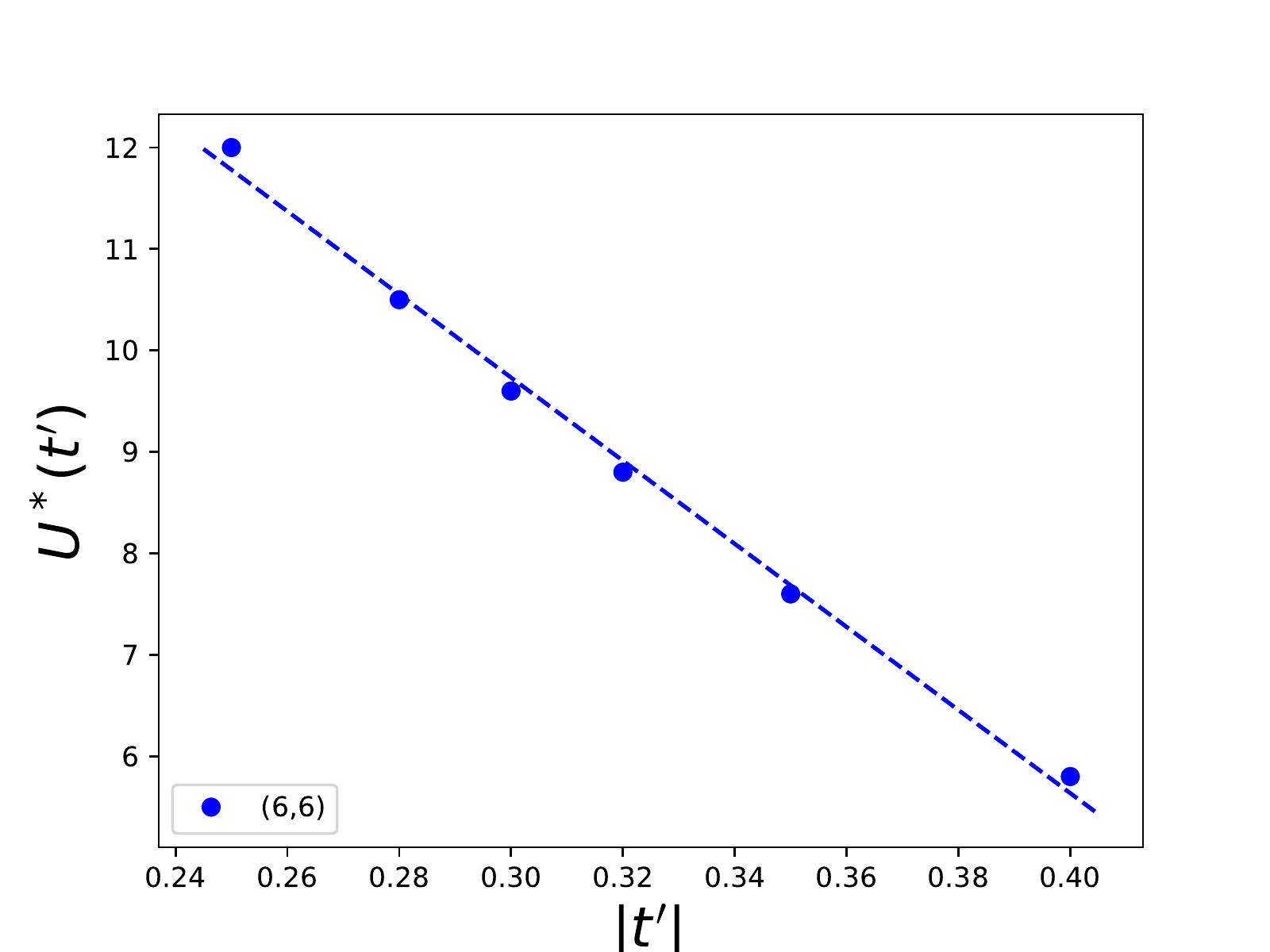}
\caption{Dependence of the critical Coulomb repulsion $U^*$ on the next-neighbor hopping $t'$, as the latter is varied in the range $t' \in \left[-0.4,\,-0.25 \right]$ for periodic boundary conditions at inverse temperature $\beta = 100$, sector $(6,6)$. The points correspond to locations of disparity minimum.  \label{fig:complex-2} 
}
\end{figure}

\end{document}